\documentclass[useAMS,usenatbib]{mnras}

\usepackage{amsmath}
\usepackage{amssymb}
\usepackage{graphicx}
\usepackage{times}

\title[Alternative method to generate pre-ICs]
  {An alternative method to generate pre-initial conditions for cosmological $N$-body simulations}

\author[Liao]
{Shihong Liao$^{1}$ \thanks{Email: liaoshong@gmail.com}
\\
$^1$Key Laboratory for Computational Astrophysics, National
Astronomical Observatories, Chinese Academy of Sciences, Beijing,
100012, China\\
}
\begin{document}

%\date{Accepted 20xx xxx xx. Received 20xx xxx xx; in original form 201x Xxx xx}

%\pagerange{\pageref{firstpage}--\pageref{lastpage}} \pubyear{2017}

\maketitle

\label{firstpage}

\begin{abstract}
Currently, grid and glass methods are the two most popular choices to generate uniform particle distributions (i.e., pre-initial conditions) for cosmological $N$-body simulations. In this article, we introduce an alternative method called the capacity constrained Voronoi tessellation (CCVT), which originates from computer graphics. As a geometrical equilibrium state, a CCVT particle configuration satisfies two constraints: (i) the volume of the Voronoi cell associated with each particle is equal; (ii) every particle is in the center-of-mass position of its Voronoi cell. We show that the CCVT configuration is uniform and isotropic, follows perfectly the minimal power spectrum, $P(k)\propto k^4$, and is quite stable under gravitational interactions. It is natural to incorporate periodic boundary conditions during CCVT making, therefore, we can obtain a larger CCVT by tiling with a small periodic CCVT. When applying the CCVT pre-initial condition to cosmological $N$-body simulations, we show that it plays as good as grid and glass schemes. The CCVT method will be helpful in studying the numerical convergence of pre-initial conditions in cosmological simulations. It can also be used to set up pre-initial conditions in smoothed-particle hydrodynamics simulations.
\end{abstract}


\section{Introduction}\label{sec_intro}
Structure formation is a fundamental problem in physical cosmology. Due to the long range and non-linear nature of gravity, currently the only known viable solution is numerical simulations. In the past decades, cosmological $N$-body simulations have deepened significantly our understanding in the formation of large-scale structures in our Universe, and contributed importantly to the establishment of the standard cosmological model, the cosmological constant $\Lambda$ $+$ cold dark matter ($\Lambda$CDM) model \citep[see e.g.][for a review]{frenk2012}.

To perform a cosmological $N$-body simulation, the very first step is to prepare an initial condition. To generate an initial condition, we need to set up a uniform distribution of $N$ particles before applying the Lagrangian perturbation theory to perturb their positions and velocities according to a given matter power spectrum. Such a uniform particle distribution is called a \textit{pre-initial condition} \citep[see e.g.,][]{baertschiger2002, hansen2007, joyce2009} or a \textit{particle load} \citep[see e.g.,][]{white1996, wang2007, jenkins2010}. A pre-initial condition should contain as less power as possible, so that it does not surpass the given physical power spectrum on the simulated scales. Therefore, a random particle distribution is not adopted because it has Poisson noise present on all scales. Currently, the known methods to generate pre-initial conditions are grid \citep[e.g.][]{efstathiou1985}, glass \citep[][]{white1996}, and Quaquaversal tiling \citep[or `Q-set',][]{hansen2007}.

The grid pre-initial condition simply places $N$ particles in the grid points of a three dimensional lattice. It is perfectly uniform and efficient to produce (i.e., its computational complexity is $O(N)$). It has no power on scales larger than the mean inter-particle separation. However, it contains unnatural lattice features, and such features persist in low-density regions (e.g., voids) of simulations at redshift $z=0$.

The glass pre-initial condition evolves a random configuration of $N$ particles under anti-gravity until the configuration reaches an equilibrium state. Since there is no preferred direction for gravity (or anti-gravity), the glass configuration is expected to be uniform and isotropic. The power spectrum of a glass configuration is close to the theoretical minimal power spectrum, $P(k)\propto k^4$ \citep[][]{zeldovich1965, peebles1980, baugh1995, wang2007}. 
The computational complexity of glasses is the same as that of $N$-body simulations. Although the glass configuration is an ideal pre-initial condition in concept, it is not easy to produce a good glass in practice. As pointed out by \citet[][]{wang2007}, the quality of a glass configuration can be affected by the gravity solver.

The Q-set pre-initial condition uses the quaquaversal tiling \citep[][]{conway1998} to recursively partition the space into a desired level, and places a particle in each quaquaversal tile. The Q-set configuration contains $N=2\times 8^{N_Q}$ particles in a rectangular box with sides of length $1$, $1$ and $\sqrt{3}$ after $N_Q$ steps, since a parent quaquaversal tile is decomposed into eight child tiles in each step. Then one has to extract a cubic region from this rectangular configuration in order to perform a simulation with a cubic box. This might cause some problems in the boundaries if one uses periodic boundary conditions. \citet[][]{wang2007} showed that comparing to grids and glasses, the Q-set method produces much more artificial structures in hot dark matter simulations.

Currently, the popular choices to prepare pre-initial conditions are grid and glass methods. Glass is the only known method to produce a uniform and isotropic pre-initial condition, and it can be affected by the gravity solver used to compute it. Unfortunately, we do not have any analytical ways to address such effects on the final simulated structures. Therefore, an interesting question is whether there are any other alternative and independent methods. If yes, the alternatives will be very helpful for us to understand the impacts of pre-initial conditions on the simulated results. Understanding and quantifying such percent-level impacts from numerical methods is a demand in the era of precision cosmology.

Here, we introduce an alternative method to set up pre-initial conditions, the capacity constrained Voronoi tessellation \citep[CCVT,][]{balzer2009}, which originates from computer graphics. The details of CCVTs will be provided in Section \ref{sec_met}. In computer graphics, generating a uniform and isotropic point distribution (also called a ``blue noise'' distribution) is an essential task. The CCVT is an algorithm put forward recently to produce high-quality blue noise point sets. We also notice that algorithms similar to the glass method were proposed recently in computer graphics to generate blue noise configurations \citep[][]{wong2017}.

This paper is structured as follows. In Section \ref{sec_met}, we introduce the CCVT algorithm and our computer code. In Section \ref{sec_prop}, we study the properties of CCVT particle configurations (e.g., power spectrum and stability under gravitational interactions). In Section \ref{sec_app}, we apply CCVT pre-initial conditions in cosmological $N$-body simulations of the $\Lambda$CDM model. Our conclusions are summarized in Section \ref{sec_con}.

\section{Methods} \label{sec_met}
Given $N_p$ particles in an $n$-dimensional space, Voronoi tessellations (VTs) are a mathematical way to partition space into $N_p$ sub-spaces, i.e., Voronoi cells. For a particle $p_i$, its associated Voronoi cell consists of all spatial points that are closer to $p_i$ than other particles $p_j$ (here $j \neq i$). We refer the reader to \citet[][]{okabe2000} for an overview of VTs. Recently, VTs see an increase in applications in observational and numerical cosmology, for example, estimating density from a discrete sample \citep[][]{schaap2000, vandeweygaert2009}, finding clusters in galaxy surveys \citep[e.g.,][]{ramella2001, kim2002, lopes2004, soares2011}, identifying halos \citep[][]{neyrinck2005} and voids \citep[][]{platen2007, neyrinck2008} from simulations, modelling of strong gravitational lenses \citep[e.g.,][]{vegetti2009, nightingale2015}, the moving mesh method for hydrodynamic simulations \citep[][]{springel2010}, etc.

There is a famous subgroup of VTs named centroid Voronoi tessellations \citep[CVTs, see][for an overview]{du1999}, whose particles coincide with the centroids (i.e., centers of mass) of Voronoi cells. The CVT can be realized with Lloyd's algorithm \citep[][]{lloyd1982}, and it has a wide range of applications. For example, in the AREPO code, CVTs are adopted to reduce mesh irregularity \citep[see][for details]{springel2010}.

The CCVT adopted in this work is a variant of CVTs, and it satisfies two constraints: (i) every particle has equal \textit{capacity} (or every Voronoi cell has equal volume), and (ii) every particle is in the center-of-mass position of its associated Voronoi cell. Intuitively, the constraint of equal volume implies that all particles have the same importance, whereas the constraint of centroidal locations means that all particles try to stay away from each other. In addition, there is no implication of preferred directions from these constraints. Therefore, we expect that a CCVT particle distribution should be uniform and isotropic. A CCVT particle distribution can be generated with the method outlined in \citet[][]{balzer2009}, which combines the algorithm described in \citet[][]{balzer2008} and Lloyd's algorithm.

\subsection{CCVT algorithm} \label{subsec_alg}

\begin{figure*} 
\centering\includegraphics[width=500pt]{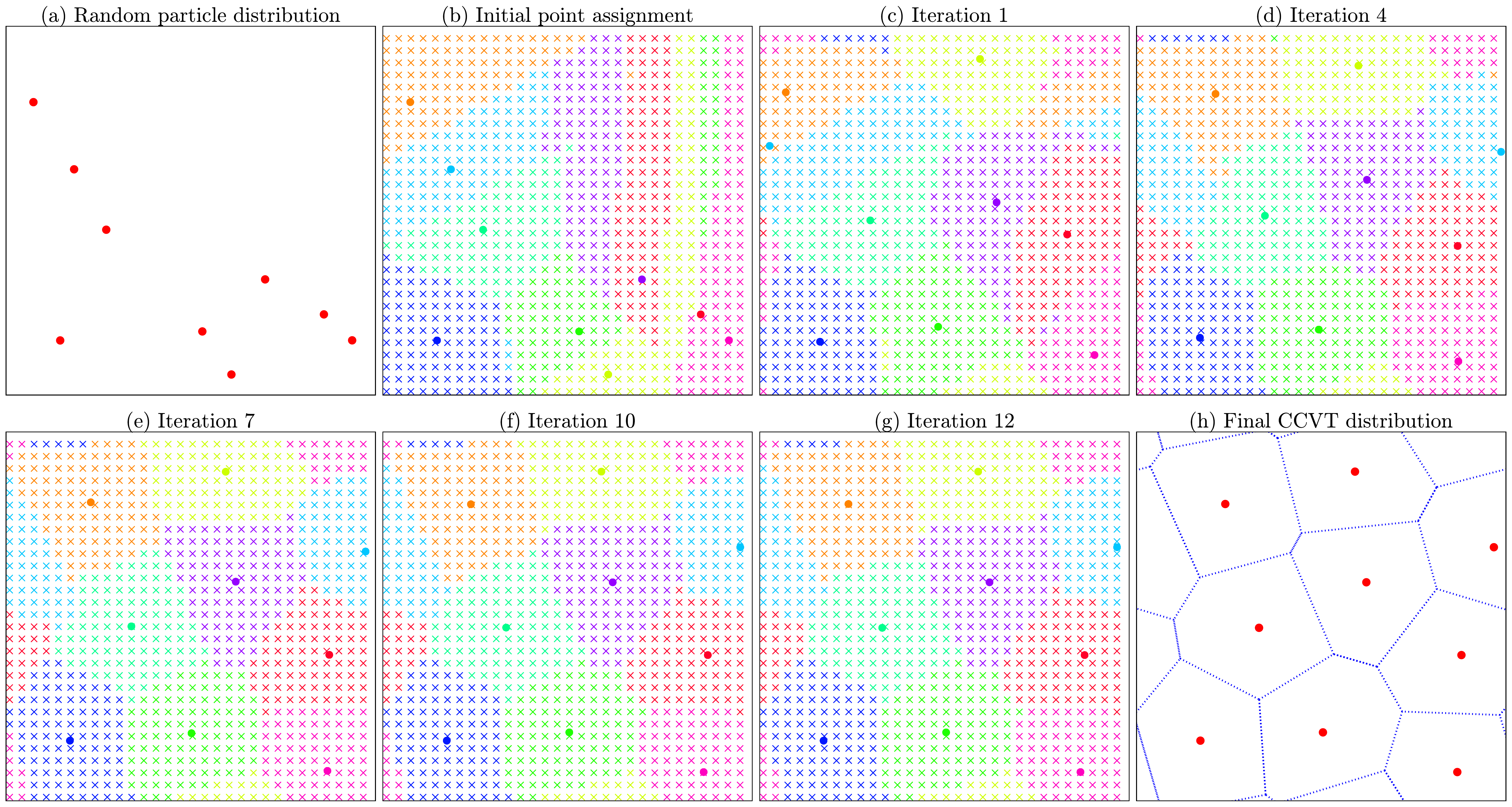} 
\caption{Illustration of the relaxation process from an initial random particle distribution to a final CCVT distribution. (a) A random distribution of 9 particles in a periodic square. (b) The initial assignment of spatial points to particles. Here we use 900 spatial points (crosses) to uniformly sample the square, and assign 100 spatial points to each particle (dots). The assignment function is shown with colors, i.e., spatial points are assigned to the particle with the same color. (c-g) Distributions of particles and their assigned spatial points after 1, 4, 7, 10, and 12 iterations. Note that, in iteration 12, there is no swapping of assignments in any particle pair, and the relaxation process is terminated. (h) Final CCVT distribution. Its VT is shown with dotted lines.}\label{fig_iter_visua}
\end{figure*}

In the method of \citet[][]{balzer2009}, we start from a random distribution of $N_p$ particles in the targeted region $R$, and then iteratively relax it to a CCVT distribution. Here we use an example of $N_p=9$ particles in a 2-dimensional periodic square to illustrate this relaxation process, as shown in Figure \ref{fig_iter_visua}. The initial random distribution is plotted in Panel (a) of Figure \ref{fig_iter_visua}.

To satisfy the first constraint of the CCVT distribution, we use $N_s$ spatial points to uniformly sample the region $R$, and assign $c=N_s/N_p$ of them to each particle; see Panel (b) of Figure \ref{fig_iter_visua} for the example with $N_s=900$ and $c=100$. By doing so, we create an assignment $A$ that each particle possesses the same number of spatial points (i.e., every particle has equal capacity $c$). In the following relaxation process, we keep the capacity of each particle fixed, and only change the detailed assignment. Note that, in the initial assignment, we can randomly assign spatial points to particles, as long as every particle gets the same capacity. But to accelerate the relaxation, an initially compact assignment according to the particle - spatial point distance is preferred, as the case shown in Panel (b) of Figure \ref{fig_iter_visua}.

It has been proved that the CCVT distribution corresponds to the minimum of the energy function \citep[][]{aurenhammer1998, balzer2008}
\begin{equation}
E=\sum_{i \in N_s} \left|\bmath{x}_i - \bmath{r}_{j=A(i)} \right|^2,
\end{equation}
where $\bmath{x}_i$ and $\bmath{r}_j$ are the coordinates of the $i$-th spatial point and the $j$-th particle respectively, and the assignment function $A(i)$ tells us the index of the particle that the $i$-th spatial point is assigned to. Therefore, the relaxation process is equivalent to minimizing the energy function of the particle - spatial point system. Note that when computing the energy function, we can easily incorporate the periodic boundary condition, and thus the final CCVT configuration is periodic and can be naturally used in cosmological simulations.

In each iteration, we loop over every pair of particles, and for each particle pair, we check over every two spatial points that belong to different particles. If the swap of the assignments of two spatial points can reduce the energy function, we swap their assignments; if not, we do nothing and move on. Note that the swapping does not modify particles' capacities. At the end of an iteration, we get a new assignment $A$ that moves the energy function closer to its minimum. After the checking and swapping operations in each iteration, similar to Lloyd's algorithm, we move every particle to the center-of-mass position of its associated spatial points so that the distribution follows the second constraint. Then we start another iteration, until there is no further swapping of assignments. See Panels (c-g) of Figure \ref{fig_iter_visua} for the particle - spatial point system after some iterations. Once there is no swapping in an iteration, the program is terminated. The final particle distribution is the desired CCVT distribution (Panel (h) of Figure \ref{fig_iter_visua}).

The worst time complexity in each iteration of this algorithm is $O(N_p^2(1+c\log c))$, where $N_p^2$ comes from that we need to loop over all particle pairs in an iteration, and $c\log c$ originates from the fact that we need to check over the spatial point pairs for each particle pair. We can adopt a heap data structure or a sorting operation to accelerate the checking and swapping process, and this brings us a complexity of $c\log c$; see \citet[][]{balzer2008} for details. Usually, since $c \gg 1$, the worst time complexity is approximated to $O(N_p^2 c\log c)$. Due to this challenging computational complexity, many efforts have been made to accelerate the computations \citep[e.g.,][]{balzer2008, li2010, ahmed2017}.

In the CCVT algorithm, the capacity is an important parameter which determines the precision of the final particle configuration and the computational time. According to the tests in Appendix \ref{ap_capacity}, the capacity is usually set to a value between $10^3$ and $20^3$. Unless specified, for the CCVT particle configurations shown in the remainder of this paper, we adopt $c=20^3$.

Note that we consider periodic boundary conditions in the computation of CCVTs, therefore, similar to glass pre-initial conditions, we can obtain a larger CCVT by tiling from a small CCVT. In Appendix \ref{ap_tile}, we test the validity of tiling CCVTs by applying them in cosmological $N$-body simulations. Our results show that the final power spectrum from a tiling CCVT pre-initial condition differs from the one without tilings only at a level of $\la 2 \%$.

\subsection{Code realization} \label{subsec_code}
We provide a code to realize the CCVT algorithm outlined in the previous subsection. The code is written in C, adopts the \textsc{Qhull} library\footnote{http://www.qhull.org} \citep[][]{barber1996} to perform Voronoi/Delaunay tessellations, uses OpenMP for parallelization, and is publicly available at https://github.com/liaoshong/ccvt-preic.

\begin{figure} 
\includegraphics[width=245pt]{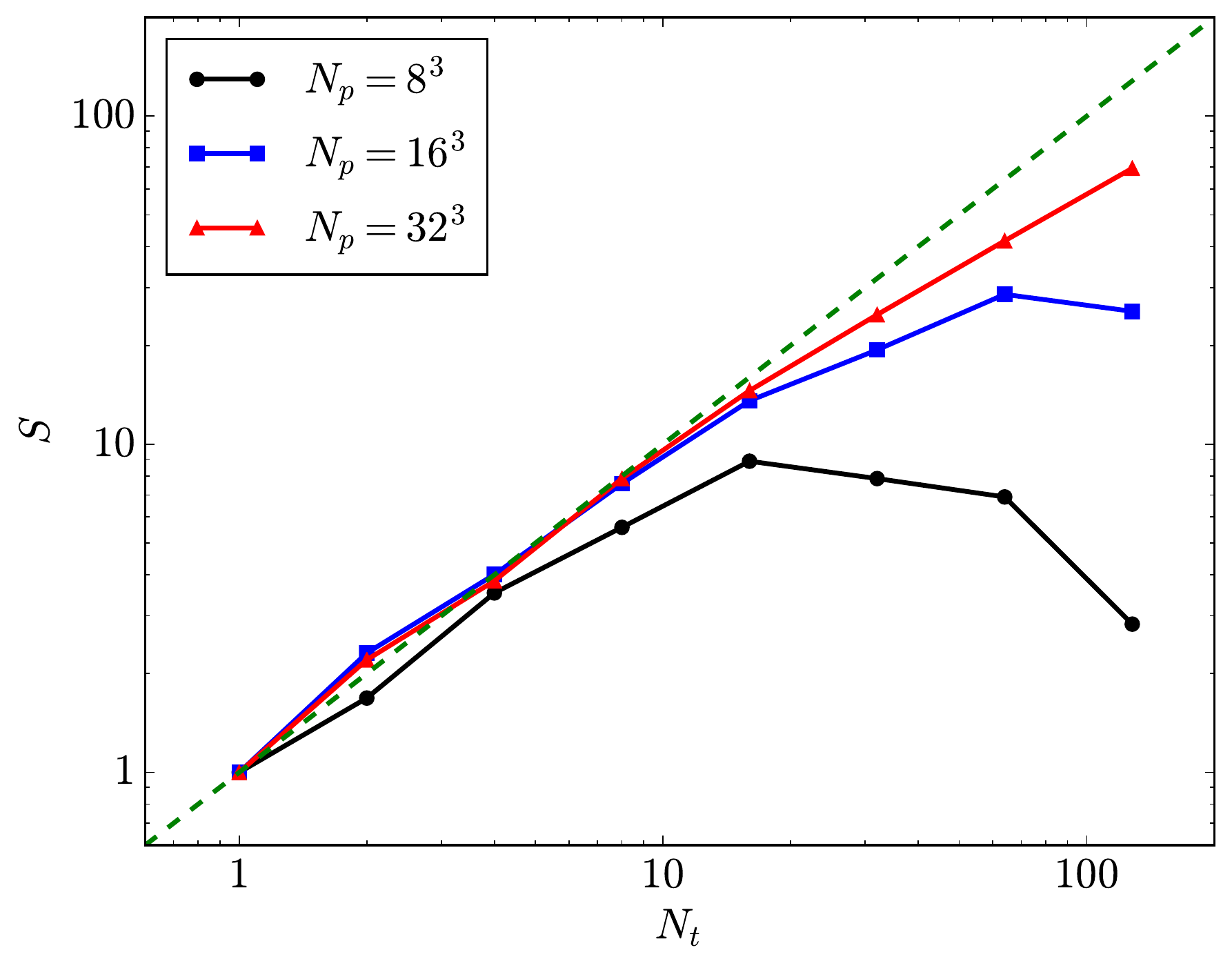} 
\caption{Scalability properties of OpenMP parallelizations. The speedup factors $S$ are plotted as a function of thread number $N_t$. The black, blue and red curves show the cases of CCVT configurations with $N_p=8^3, 16^3$ and $32^3$ particles, respectively. The green dashed line shows a linear speedup. All CCVTs shown in this figure are computed with $c=10^3$.}\label{fig_scaling}
\end{figure}

To accelerate the code, we use OpenMP to parallelize the checking and swapping computations in each iteration. When the code runs with $N_t$ threads, it divides $N_p$ particles into $2N_t$ subgroups. The computations can be classified into two types, i.e., (i) computing particle pairs whose particles belong to the same subgroup, and (ii) computing particle pairs whose particles belong to different subgroups. For type (i), it is simple that each thread is responsible to two subgroups and computes independently. For type (ii), there are $N_t(2N_t - 1)$ pairs of subgroups in total. In each round, each thread considers one pair of subgroups, and the computations are finished in $(2N_t - 1)$ rounds. Note that in each round, to avoid race conditions, a subgroup can only be accessed by one thread. Under such requirements, how to assign subgroup pairs to different threads in different rounds is a problem of balanced tournament designs in combinatorial mathematics, and a solution can be found in \citet[][]{colbourn2007}. The parallelization scaling of our code is presented in Figure \ref{fig_scaling}, where we plot the speedup, $S\equiv T_1/T_{N_t}$, as a function of thread number, $N_t$. Here, $T_{N_t}$ is the wall time for the code running with $N_t$ threads. For a small problem size with too many threads (e.g., $N_p=8^3$ with $N_t>16$), it becomes harder to balance the work equally among the threads, and the code spends more time on waiting. Apart from such cases, our code exhibits relatively good scalability properties. 

The computations can also be accelerated if we start from a quasi-uniform and isotropic particle distribution instead of a random one. Therefore, we provide an option in the code that it uses the capacity constrained Delaunay triangulation \citep[CCDT,][]{xu2011} to relax the random distribution for several steps before starting CCVT iterations. CCDT is an efficient method to obtain a blue noise particle distribution in computer graphics. By evolving a random distribution with CCDT for several steps, we get a quasi-uniform and isotropic particle distribution, which is closer to the minimum of the energy function. For example, if we solve a problem of $N_p=16^3$ and $c=10^3$ with one thread using an Intel Core i7-4790 processor, it takes $755.95$ seconds (i.e., $2.64$ seconds for initializations $+$ $751.37$ seconds for CCVT relaxations $+$ $1.94$ seconds for other operations) measured in CPU time without CCDT initializations. However, with CCDT initializations, it only takes $421.71$ seconds (i.e., $10.18$ seconds for initializations $+$ $409.35$ seconds for CCVT relaxations $+$ $2.18$ seconds for other operations), which results in a speedup factor of $\sim 1.8$.

\section{Properties}\label{sec_prop}
\begin{figure*} 
\centering\includegraphics[width=500pt]{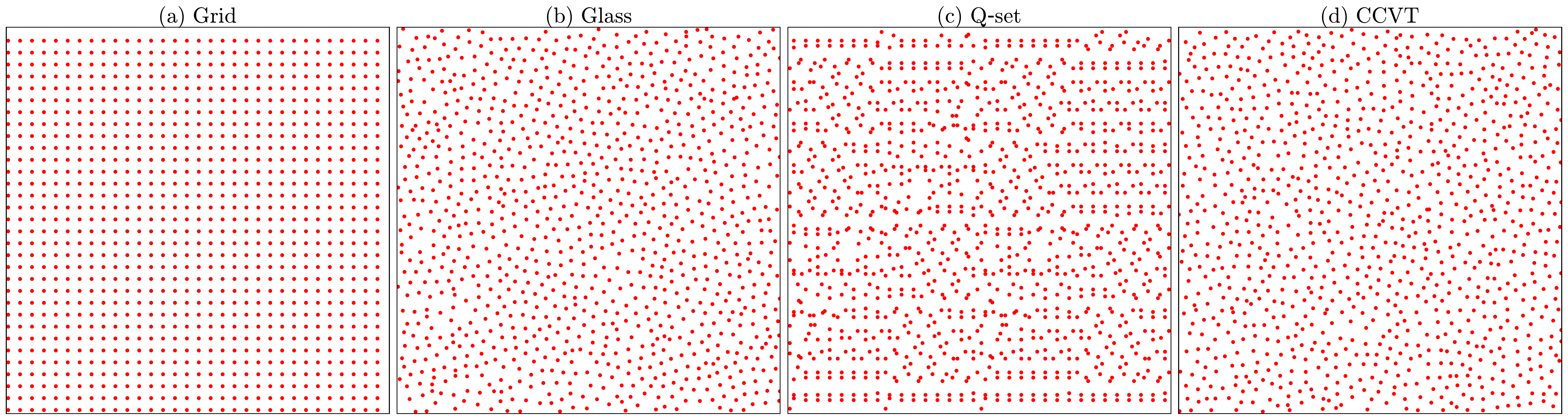} 
\caption{Visualizations of grid, glass, Q-set and CCVT particle distributions. For each panel, we select a slice with a thickness of a mean particle separation along the $z$-axis from a box of $32^3$ particles, and project it into the $xy$-plane. Each panel contains roughly $32^2$ particles. The Q-set particles are obtained with $N_Q = 5$.}\label{fig_preic_visua}
\end{figure*}

In Figure \ref{fig_preic_visua}, we present a visualization of grid, glass, Q-set and CCVT particle distributions. In this study, we use the hybrid Tree-Particle Mesh (TreePM) gravity solver, \textsc{Gadget-2}\footnote{http://www.mpa-garching.mpg.de/gadget} \citep[][]{springel2005}, to generate glass distributions. Similar to the glass case, the CCVT distribution possesses a fairly good quality in uniformity and isotropy. In contrast, the Q-set particles contain some characteristics of lattice effects \citep[see also the discussions in][]{diehl2015}. We further quantify their properties in the following subsections.

\subsection{Power spectra}

\begin{figure} 
\includegraphics[width=245pt]{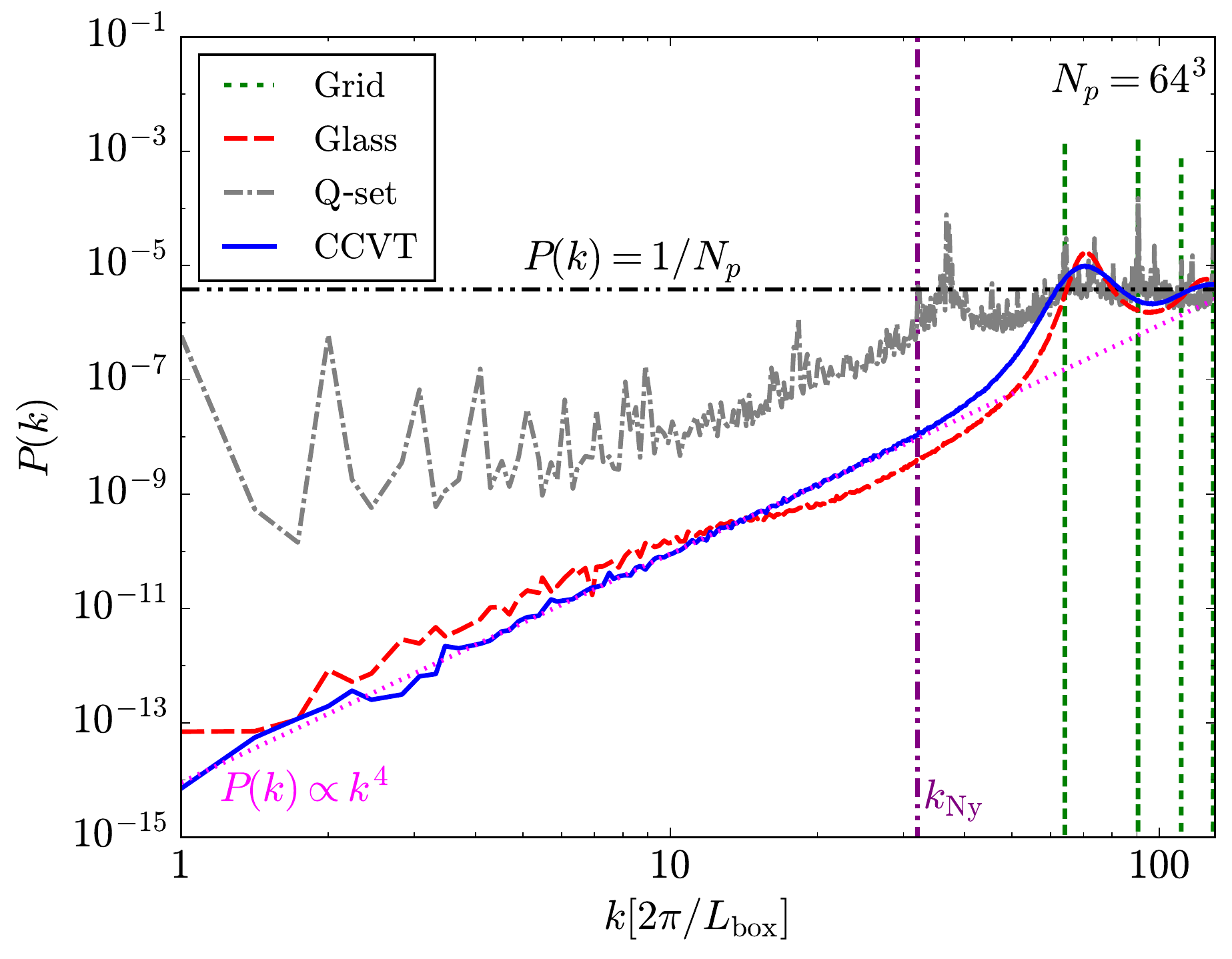} 
\caption{Power spectra of different pre-initial conditions with $64^3$ particles. The power spectra of grid, glass, Q-set and CCVT distributions are shown with green short-dashed, red long-dashed, grey dash-dotted and blue solid lines, respectively. The Poisson noise power spectrum is indicated with a dash-dot-dotted horizontal line, and the $k^4$-minimal power spectrum is plotted with a magenta dotted line. The Nyquist frequency is marked as a dash-dot-dotted vertical line. We use very fine binnings in $k$ to compute $P(k)$ here.}\label{fig_powspec}
\end{figure}

The power spectrum, $P(k)$, of a CCVT configuration with $64^3$ particles is given in Figure \ref{fig_powspec}, together with power spectra for other pre-initial conditions. Here, $P(k)$ is defined as
\begin{equation}
\left<\tilde{\delta}(\bmath{k}) \tilde{\delta}^\ast (\bmath{k}^\prime)\right> = \left(2\pi \right)^3 P(k)\delta^D(\bmath{k} - \bmath{k}^\prime),
\end{equation}
where $\tilde{\delta}(\bmath{k})$ is the Fourier transform of $\delta(\bmath{x})\equiv \rho(\bmath{x})/\bar{\rho} - 1$, $\rho(\bmath{x})$ is the density at position $\bmath{x}$, $\bar{\rho}$ is the global mean density, and $\delta^D(\bmath{x})$ is the 3-dimensional Dirac delta function. To compute $P(k)$, we assign particle masses to a density field in a mesh with the Cloud-in-Cell (CIC) scheme, and apply the FFTW library\footnote{http://fftw.org/} \citep[][]{frigo2005} to perform Fourier transforms.

\citet[][]{zeldovich1965} and \citet[][]{peebles1980} showed that for a uniform discrete particle distribution, the minimal power spectrum is $P(k) \sim k^4$. From Figure \ref{fig_powspec}, we can see that the CCVT pre-initial condition tightly follows this $k^4$-power law up to the Nyquist frequency, $k_\mathrm{Ny}$. This indicates that the CCVT configuration is indeed uniform. Interestingly, it is even closer to this $k^4$-power law relation than the glass one which deviates a little from the $k^4$-relation at $k\sim 10$ \citep[see also][]{wang2007, joyce2009}. At small scales ($k \ga 64$), the spectra are taken over by Poisson noise, $P(k) = 1/N_p$. The differences in the power spectra of CCVT and glass configurations indicate that they are two different particle distributions. The CCVT distribution represents a geometrical equilibrium while the glass distribution is a dynamical equilibrium.

The grid configuration only has non-zero power in typical inter-particle separations. The Q-set power spectrum exhibits many spikes, which arise from the characteristic inter-particle distances \citep[][]{hansen2007}. Such grid-like features are also common in other tiling methods for blue noise sampling in computer graphics \citep[see][for examples]{lagae2008}. The Q-set power spectrum is not a perfect $k^4$-relation, but only approximates to it \citep[see also][]{wang2007}. In the remainder of this article, we mainly compare the CCVT pre-initial condition with the grid and glass ones.

\subsection{Force balance}
One shared property for grid and glass configurations is force balance, i.e., each particle feels zero gravitational force from other particles. Therefore, before adding perturbations with the Lagrangian perturbation theory, they are stable under the evolution with gravitational interactions. The grid configuration achieves this by symmetry, whereas the glass configuration satisfies this by definitions. What about the case for CCVT configurations? This is a non-trivial question. Instead of offering an analytical answer to this question, in the following, we numerically evolve a CCVT configuration in a standard cold dark matter (SCDM, i.e., the matter fraction $\Omega_m=1$) universe from $a=1$ ($a$ being the scale factor) to $1000$ under gravitational forces, and see whether structures form. As a comparison, we perform a parallel analysis on a numerical realization of glass configurations.

\begin{figure*} 
\centering\includegraphics[width=500pt]{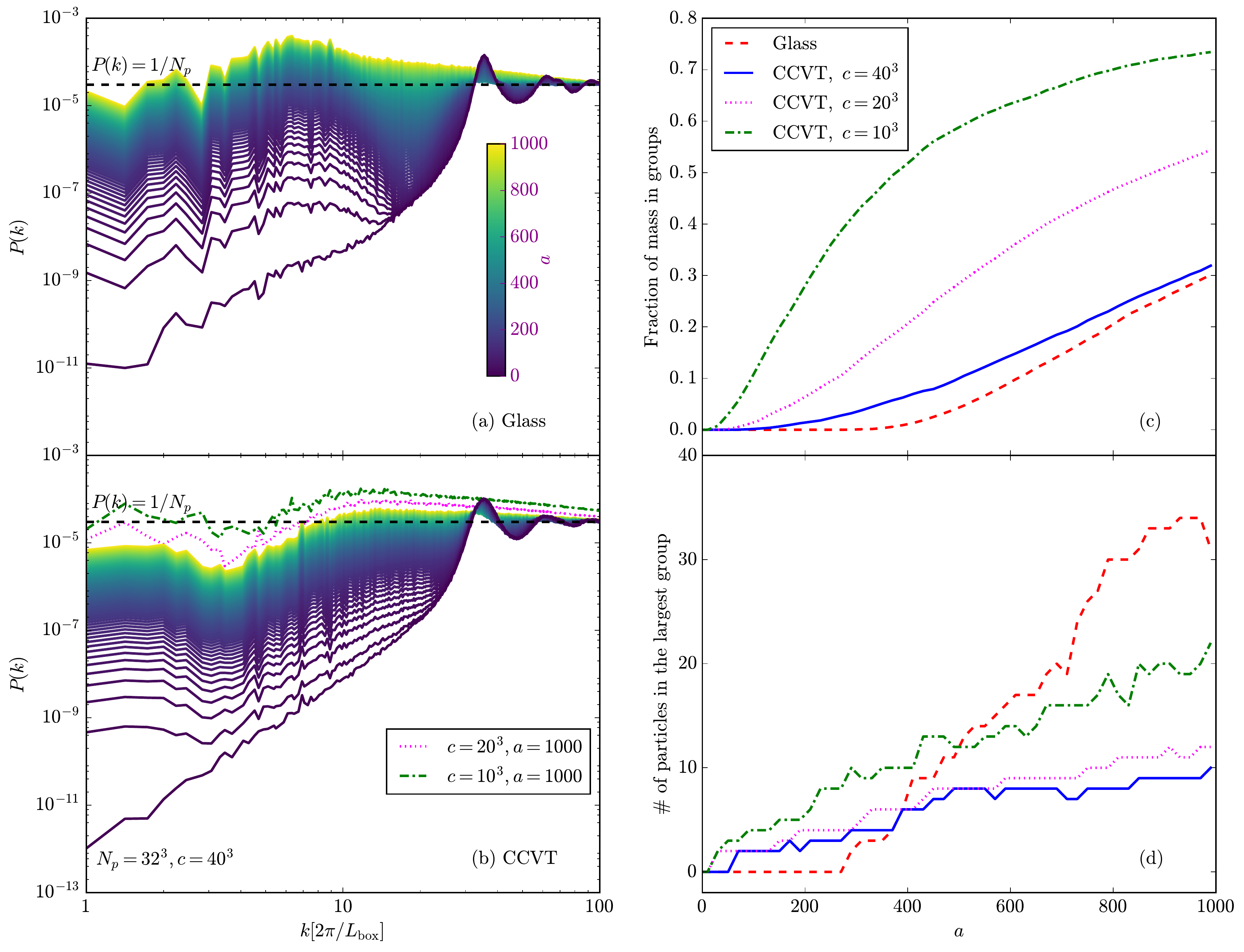} 
\caption{Evolution of glass and CCVT configurations under gravitational forces in an SCDM universe. Panels (a) and (b) present the evolution of power spectra of glass and CCVT ($c=40^3$) configurations from $a=1$ to $a=1000$, respectively. Colors indicating different scale factors are specified by the color bar in Panel (a). The black dashed horizontal lines mark the Poisson noise power spectra. As a comparison, we also plot the power spectra at $a=1000$ for the CCVT cases with $c=10^3$ (green dash-dotted) and $20^3$ (magenta dotted) in Panel (b). Panels (c) and (d) show the fraction of mass in the identified FOF groups and the number of particles in the largest FOF group at different $a$, respectively. The red dashed (blue solid) curve gives the glass (CCVT with $c=40^3$) results. Results from CCVTs with $c=10^3$ (green dash-dotted) and $20^3$ (magenta dotted) are also plotted for comparison.}\label{fig_force_balance}
\end{figure*}

The CCVT configuration used in this investigation contains $32^3$ particles, and it is generated with a capacity of $40^3$. The glass configuration is generated by \textsc{Gadget-2} with the same number of particles. We evolve these two configurations under the pure Tree gravity of \textsc{Gadget-2} with a comoving softening length equalling to $1/50$ of the mean inter-particle separation. After the evolution, we adopt the friends-of-friends \citep[FOF,][]{davis1985} algorithm with a linking length parameter of $0.2$ to identify structures which contain at least two particles.

The results are shown in Figure \ref{fig_force_balance}. In Panels (a) and (b), we compare the evolution of power spectra of these two configurations. As they evolve, their power spectra start to deviate from the minimal power spectrum. At $a=1000$, both of the CCVT and glass configurations have power spectra comparable to the Poisson noise one, which indicates that some structures have formed. A more direct look at the structure formation during the evolution can be shown in the FOF group identifications, which are summarized in Panels (c) and (d). Panel (c) plots the fraction of mass in FOF groups during the evolution, whereas Panel (d) shows the number of particles for the largest FOF group as a function of scale factor. We find that the CCVT configuration starts to form structures at $a\approx 100$, while the glass configuration is at $a\approx 300$. At $a=1000$, about $1/3$ of the mass is in FOF groups for both cases. The largest group in glass at $a=1000$ has $\sim 30$ particles, while it has $\sim 10$ particles in the CCVT case.

Note that this numerical investigation depends highly on the precision of numerical gravity solvers and the precision of the numerical realizations of pre-initial conditions. For example, if we evolve the glass configuration with the TreePM solver, different PM grids and splitting scales will cause the FOF structures to form earlier or later than the case shown in Figure \ref{fig_force_balance}. Given above that the CCVT configuration exhibits comparable results as the glass one, we conclude that CCVT configurations achieve relatively good force balance properties. An analytical investigation is still an open question. We have noticed from our numerical investigations that with a larger $c$, a numerical CCVT configuration can achieve balanced forces better (see Panels (b-d) in Figure \ref{fig_force_balance} for the results of two additional CCVT investigations with $c=10^3$ and $20^3$), therefore we anticipate that an ideal CCVT configuration should possess good force balance properties.

\section{Applications in cosmological simulations}\label{sec_app}
Given that the CCVT pre-initial condition exhibits good properties on the power spectrum and force balance, we further test its validity of applications in cosmological $N$-body simulations in this section.

We perform three $\Lambda$CDM simulations starting from grid, glass and CCVT pre-initial conditions respectively. Each of them has $256^3$ particles in a periodic cube with a box size of $L_\mathrm{box}=100$ $h^{-1}\mathrm{Mpc}$. The adopted cosmological parameters come from the Planck 2015 results \citep[][]{ade2016}, i.e., $\Omega_m = 0.3089$, $\Omega_b=0.0486$, $\Omega_\Lambda=0.6911$, $h=0.6774$, $\sigma_8=0.8159$, and $n_s=0.9667$. To address the convergence scales of these simulations, we run a higher-resolution simulation with $512^3$ particles starting from grid pre-initial conditions. We use the \textsc{2LPTic} code\footnote{http://cosmo.nyu.edu/roman/2LPT/} \citep[][]{crocce2006} to generate the initial conditions at $z_\mathrm{IC}=100$ with the same random phases. The gravitational softening lengths, $\epsilon$, are set as $1/50$ of the mean inter-particle separations. All simulations are performed with the \textsc{Gadget-2} code. The halos are identified using the Amiga Halo Finder\footnote{http://popia.ft.uam.es/AHF/Download.html} \citep[\textsc{AHF},][]{knollmann2009} with a virial overdensity parameter $\Delta_\mathrm{vir} = 200$ measured with respect to the critical density. Note that \textsc{AHF} returns both distinct halos and subhalos at the same time, and here we only consider the distinct ones.

\begin{figure*} 
\centering\includegraphics[width=505pt]{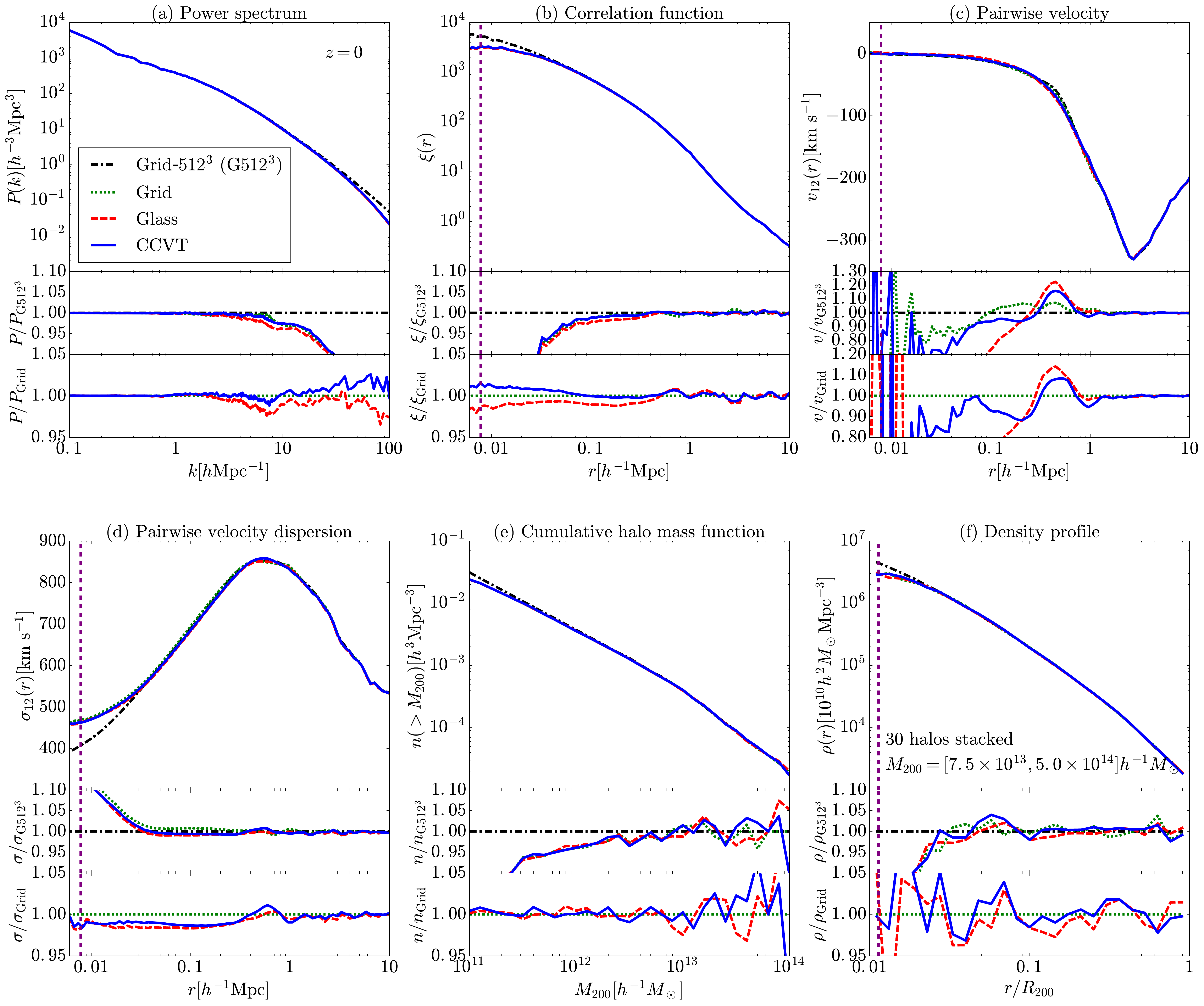} 
\caption{Comparisons of $z=0$ properties of simulations starting from different pre-initial conditions. From Panel (a) to (f), we plot their power spectra, correlation functions, pairwise velocities, pairwise velocity dispersions, cumulative halo mass functions, and stacked density profiles respectively. In each panel, we use black dash-dotted lines to plot the results from the simulation with higher resolutions, and green dotted, red dashed and blue solid lines to show the results from lower-resolution simulations starting from grid, glass and CCVT pre-initial conditions respectively. In each panel, the first row shows the physical quantities, the second row shows their ratios with respect to that of the higher-resolution simulation, and the third row shows their ratios with respect to that of the grid low-resolution simulation. The purple dashed vertical lines in Panels (b-d) mark the softening length of lower-resolution simulations, $\epsilon_{256^3}$, whereas the purple dashed line in Panel (f) marks the value of $\epsilon_{256^3}/R_\mathrm{200,min}$ (with $R_\mathrm{200,min}$ being the minimum virial radius of the stacked halo sample).}\label{fig_app_sim}
\end{figure*}

In Figure \ref{fig_app_sim}, we compare the $z=0$ properties (including the spatial and velocity information of dark matter distributions, and the global abundance and internal structures of dark matter halos) of simulations starting from different pre-initial conditions. The results are discussed as follows.

(a) Power spectra. The power spectra of all four simulations are plotted in Panel (a) of Figure \ref{fig_app_sim}. At $k \la 30$ $h\mathrm{Mpc}^{-1}$, $P(k)$ from low-resolution simulations converge to that from the high-resolution simulation at a $10\%$ level. Within this convergent region of low-resolution simulations, the difference between $P(k)$ from different pre-initial conditions is $\la 3\%$.

(b) Two-point correlation functions. The two-point correlation functions of simulated dark matter particles, $\xi(r)$, shown in Panel (b) of Figure \ref{fig_app_sim} are computed with the estimator proposed in \citet[][]{landy1993}. $\xi(r)$ from low-resolution simulations converge to that of the high-resolution simulation at a $10\%$ level for $r \ga 4\epsilon_{256^3}$, where $\epsilon_{256^3}$ is the softening length of low-resolution simulations. Within the convergent regime, $\xi(r)$ from simulations with different pre-initial conditions deviate from each other at a level of $\la 2\%$.

(c) Pairwise velocities. In Panel (c) of Figure \ref{fig_app_sim}, we present the pairwise velocities, $v_{12}(r)$, of dark matter particles from different simulations. Here, the pairwise velocity along the line-of-separation is defined as
\begin{equation}
v_{12}(r)\equiv\left<\left[\bmath{v}_1(\bmath{r}^\prime) - \bmath{v}_2(\bmath{r}^\prime+\bmath{r})\right]\cdot \hat{\bmath{r}} \right>,
\end{equation}
where $\bmath{v}_i(\bmath{r}^\prime)$ is the peculiar velocity of particle $i$ at position $\bmath{r}^\prime$, the unit vector is $\hat{\bmath{r}} \equiv \bmath{r}/|\bmath{r}|$, and $\left<\cdots\right>$ means the average over all pairs which have a separation of $r=|\bmath{r}|$. Compared to $P(k)$ and $\xi(r)$ mentioned above, we see that pairwise velocities at small scales ($r \la 1h^{-1}\mathrm{Mpc}$) are more sensitive to pre-initial conditions. For example, at $r \approx 0.4 h^{-1}\mathrm{Mpc}$, which is quite a large scale comparing to the softening length (dashed vertical line), the difference between the grid one and the glass one is $\sim 20\%$. The difference becomes more fluctuating at smaller scales. The CCVT plays somewhat between the grid and glass pre-initial conditions. This suggests that the CCVT pre-initial condition is as valid as the grid and glass ones in studying pairwise velocities.

(d) Pairwise velocity dispersions. The pairwise velocity dispersion in the line-of-separation direction is defined as
\begin{equation}
\sigma_{12}(r) \equiv \left<\left\{\left[\bmath{v}_1(\bmath{r}^\prime) - \bmath{v}_2(\bmath{r}^\prime+\bmath{r})\right]\cdot \hat{\bmath{r}}\right\}^2\right>^{1/2}.
\end{equation}
From Panel (d) of Figure \ref{fig_app_sim}, we find that $\sigma_{12}(r)$ is not sensitive to pre-initial conditions. At the convergence regime ($r \ga 2\epsilon_{256^3}$), the difference among $\sigma_{12}(r)$ from low-resolution simulations with different pre-initial conditions is $\la 2\%$.

(e) Cumulative halo mass functions. The abundances of dark matter halos in our simulations, $n(>M_{200})$, are given in Panel (e) of Figure \ref{fig_app_sim}. The difference among low-resolution simulations is again fairly small at a wide range of halo masses. However, when approaching the mass scale of $10^{14}h^{-1}M_\odot$, the difference becomes noisier. This is due to the low number of cluster halos in our simulated boxes, which causes noticeable Poisson noise. 

(f) Halo density profiles. To reduce noise and offer a robust assessment, in each simulation, we select the first $30$ massive halos from the \textsc{AHF} distinct halo catalogue, and stack them to get the density profile. The stacked halos have masses in the range of $[7.5\times 10^{13}, 5.0\times 10^{14}]h^{-1}M_\odot$. The stacked density profiles are compared in Panel (f) of Figure \ref{fig_app_sim}. At the convergence regime ($r \ga 2\epsilon_{256^3}$), the difference between density profiles from low-resolution simulations is $\la 5\%$.

From the above examination of properties covering different aspects of cosmological $N$-body simulations, we find that the simulation with a CCVT pre-initial condition gives fairly similar results as the grid and glass ones. The difference between simulations with different pre-initial conditions are usually at percent levels. Therefore, we conclude that the CCVT pre-initial condition is valid in $\Lambda$CDM simulations.

\section{Conclusions}\label{sec_con}
In this article, we introduce an alternative method, the CCVT algorithm, which originates from computer graphics, to generate pre-initial conditions for cosmological $N$-body simulations. We show that the CCVT configuration is uniform and isotropic, and follows perfectly the minimal power spectrum, $P(k)\sim k^4$, at $k \la k_\mathrm{Ny}$. To test the property of force balance, we numerically evolve the CCVT pre-initial condition in an SCDM model under gravitational interactions, and we find that the stability of CCVTs is comparable to that of glasses. We also demonstrate the validity of applying CCVT pre-initial conditions in $\Lambda$CDM cosmological $N$-body simulations. The CCVT is the second known method to set up uniform and isotropic pre-initial conditions. It will be helpful in studying the numerical convergence of pre-initial conditions in cosmological simulations, which is important to investigate, understand and quantify in the era of precision cosmology. The CCVT algorithm can also be used to set up pre-initial conditions for smoothed-particle hydrodynamics simulations.

Although here we only apply the CCVT algorithm to generate uniform and isotropic configurations, by giving non-uniform capacities, it can generate particle configurations under any given density distributions; see \citet[][]{balzer2009} for details. It will be interesting to explore if it can be applied to prepare initial conditions for cosmological zoom-in simulations.

This study is an attempt to introduce methods in computer graphics into cosmological simulations. Apart from the CCVT algorithm, there are many other methods developed in computer graphics to produce blue noise distributions \citep[see e.g.,][for reviews]{lagae2008, yan2015}. It will be interesting to see if there are any other blue noise sampling methods that have good physical properties and can be applied to cosmological simulations.

\section*{Acknowledgements}
The author is particularly grateful for the advice of testing force balances for CCVT distributions and the constructive comments on this manuscript from Adrian Jenkins, the help in computing power spectra and comments on this manuscript from Jie Wang, and the feedbacks and suggestions on the CCVT code from Jianxiong Chen. The author also thanks Marius Cautun, M.-C. Chu, Liang Gao, Baojiu Li, Ming Li, Shi Shao, Shuangpeng Sun and Jiajun Zhang for discussions, Steen H. Hansen for providing the Q-set code, and Michael Balzer for making his serial C++ CCVT code publicly available (https://code.google.com/archive/p/ccvt/). The author is thankful to the hospitalities of the Institute for Computational Cosmology, Durham University and Department of Physics, The Chinese University of Hong Kong, where part of this work was done.

\appendix
\section{Convergence tests on capacity}\label{ap_capacity}

\begin{figure} 
\includegraphics[width=245pt]{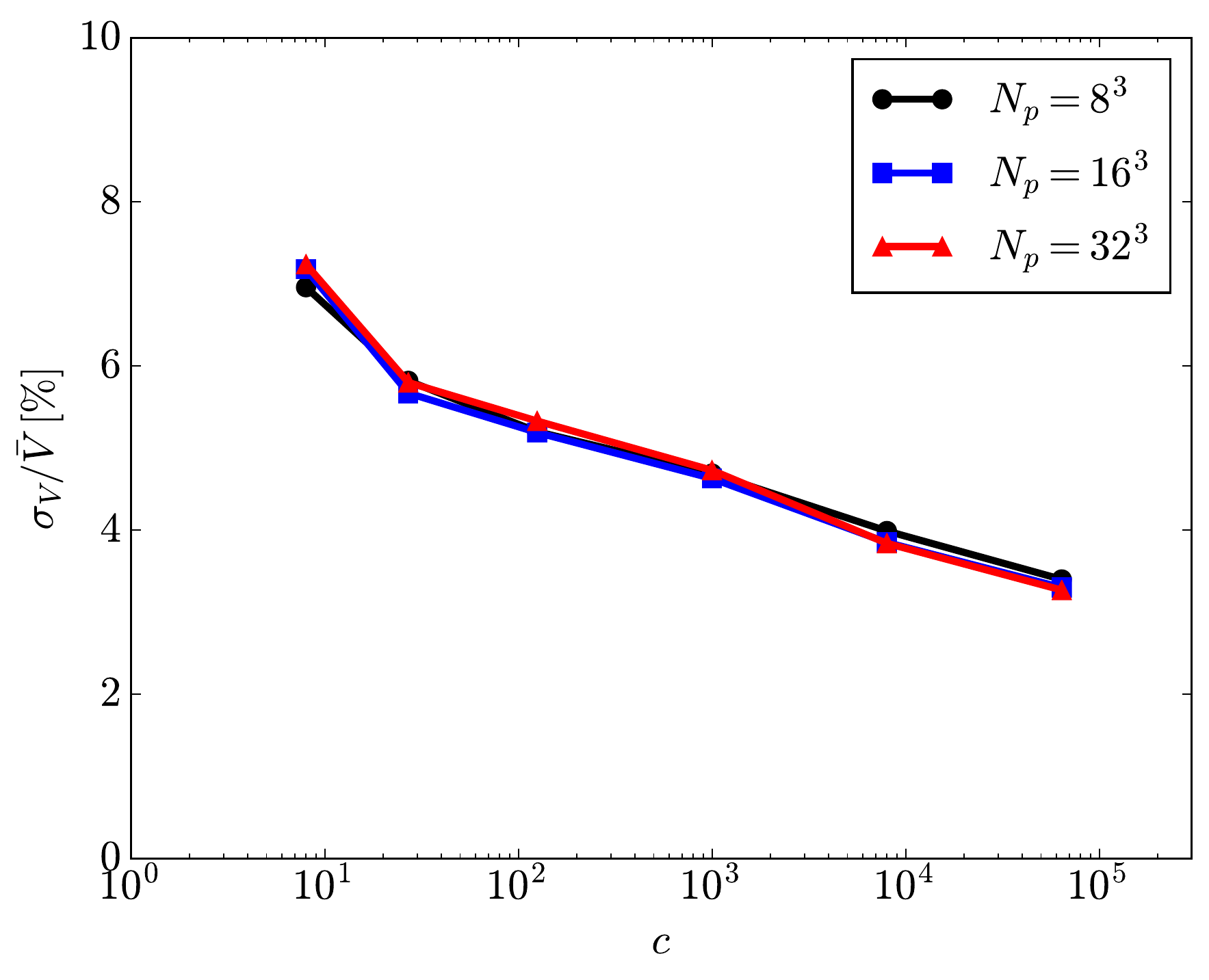} 
\caption{Relation between relative volume variance and capacity. The black, blue and red curves show the testing results from CCVT configurations with $N_p=8^3, 16^3$ and $32^3$, respectively.}\label{fig_capacity_volume}
\end{figure}

\begin{figure} 
\includegraphics[width=245pt]{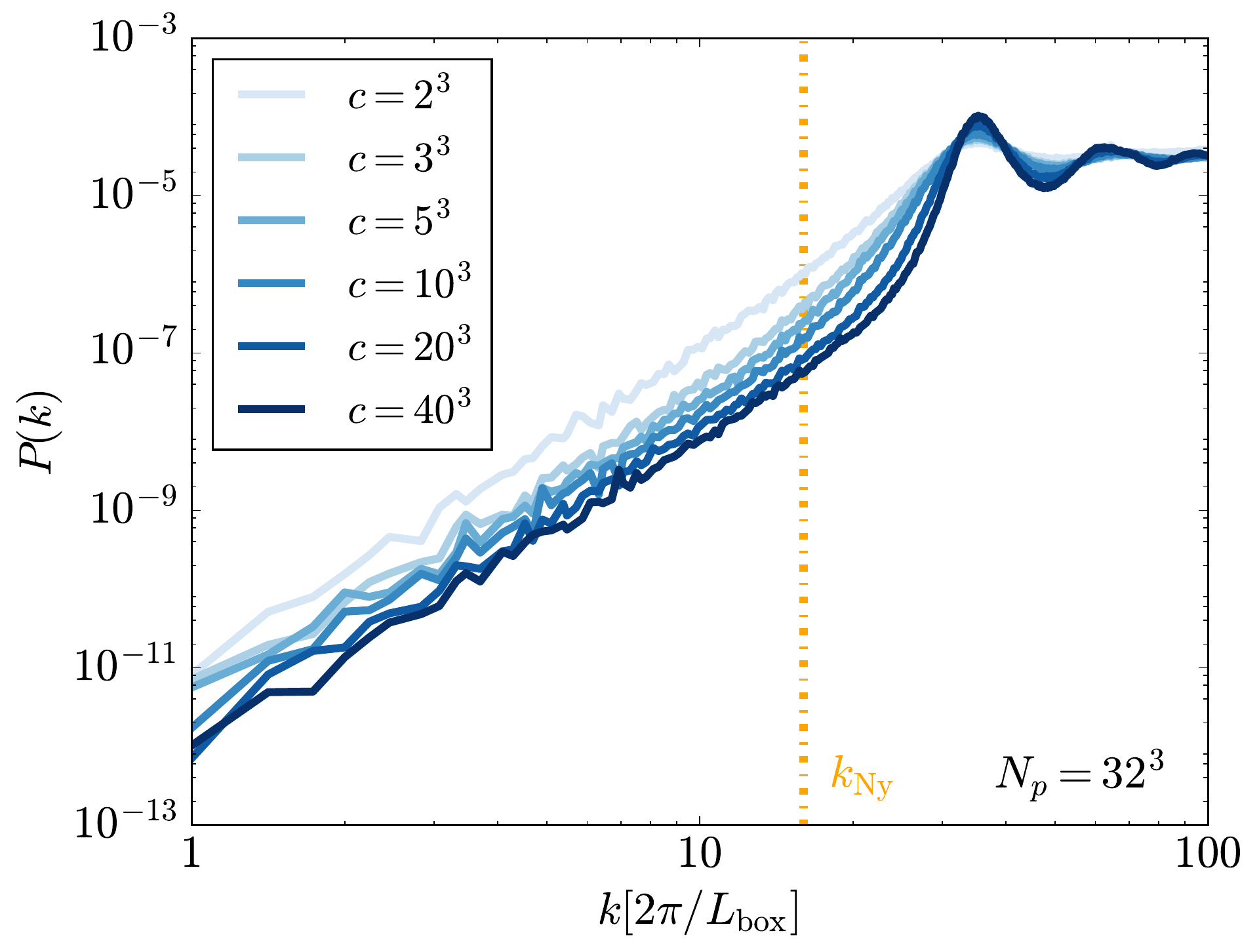} 
\caption{Power spectra of CCVT pre-initial conditions computed with different capacities. Each CCVT configuration contains $32^3$ particles, and its capacity number is given in the legend. The Nyquist frequency is marked with an orange dash-dotted vertical line.}\label{fig_capacity_powspec}
\end{figure}

\begin{figure*} 
\centering\includegraphics[width=400pt]{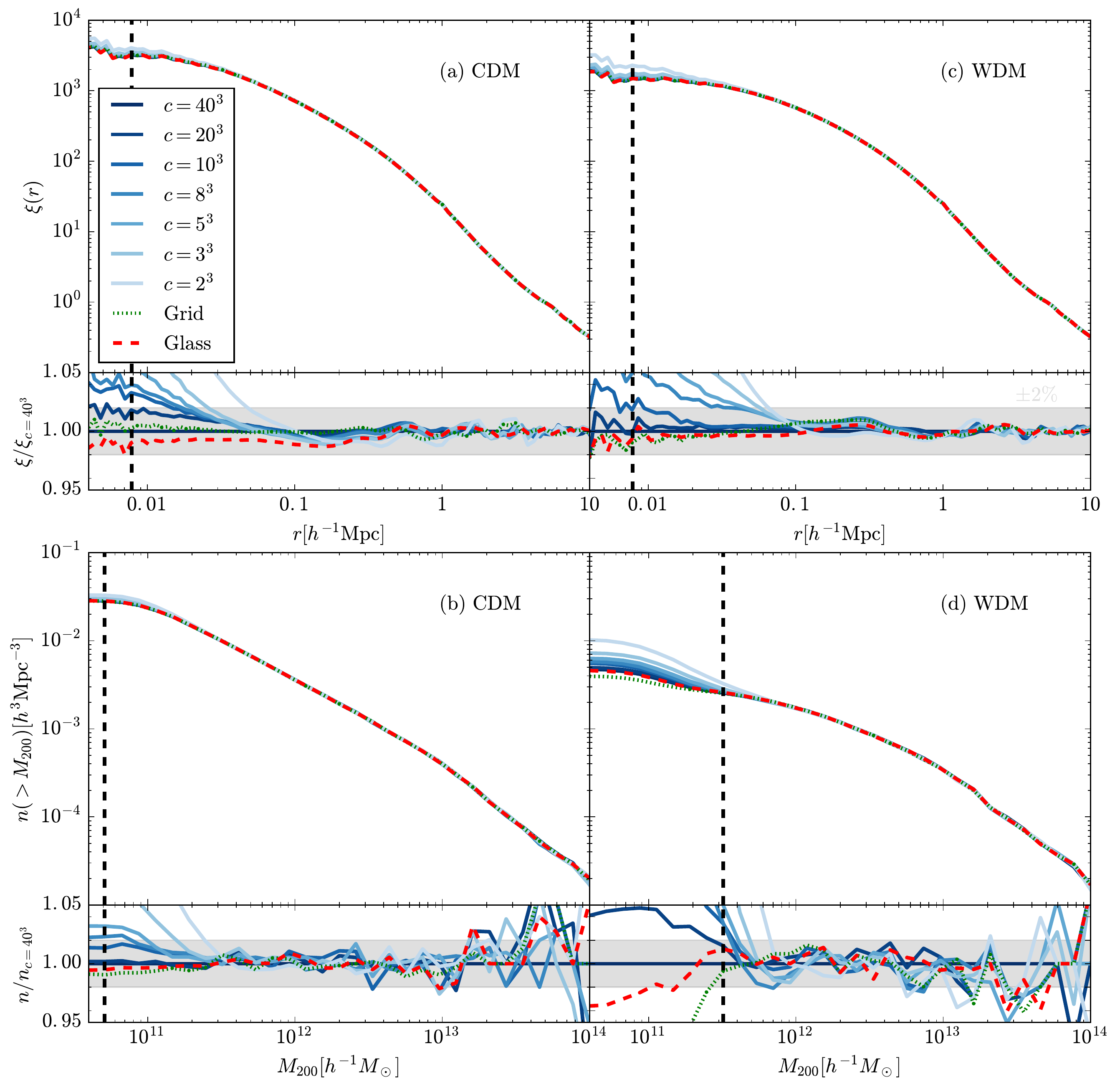} 
\caption{Convergence tests of capacities in CDM and WDM simulations. Panels (a) and (b) show the two-point correlation functions and cumulative halo mass functions at $z=0$ from CDM simulations with different pre-initial conditions, respectively. Panels (c) and (d) show the similar results for the WDM case. The colors and linestyles used to denote simulations from different pre-initial conditions are summarized in the legend of Panel (a). In each panel, the upper row shows the physical quantities, while the lower row presents their ratios with respect to the simulation starting from a CCVT pre-initial condition with $c=40^3$. In the lower row of each panel, we use a shaded region to mark the range of $1\pm 2\%$. The dashed vertical lines in Panels (a) and (c) mark the numerical softening length. In Panel (b), the dashed vertical line gives the mass of a halo with $10$ particles, whereas in Panel (d), the dashed vertical line represents the mass scale where the glass results differ from the grid ones at a level of $2\%$. Note that the noisy results for group and cluster halos (i.e., $M_{200} \ga 10^{13}h^{-1}M_\odot$) in the ratio plots of Panels (b) and (d) are due to the small number of high-mass halos in our simulations.}\label{fig_capacity_sim}
\end{figure*}

The capacity parameter affects the precision of the generated CCVT configuration and the computational time. With a larger capacity, we sample the space with higher resolution, and thus obtain a CCVT distribution which meets the constraints better. However, this will cost more computational time. To find out what values we should set for the capacity, we have performed a series of tests with different $N_p$ and $c$.

First, we look at the relative volume variance, $\sigma_V/\bar{V}$, of the final CCVT distributions, where $\bar{V}$ and $\sigma_V$ are the mean and standard deviation of the volumes of Voronoi cells respectively. A smaller relative volume variance implies a better CCVT distribution. The relation between relative volume variance and capacity is shown in Figure \ref{fig_capacity_volume}. We find that CCVTs with different $N_p$ show similar results, and to obtain a CCVT with $\sigma_V/\bar{V} \la 5\%$, we need $c \ga 10^3$.

Second, we compare the power spectra of CCVTs with $32^3$ particles obtained from different capacities. The results are plotted in Figure \ref{fig_capacity_powspec}. We find that as $c$ increases, the magnitude of the power spectrum at $k \la k_\mathrm{Ny}$ decreases, and the power spectrum with $c=20^3$ is quite close to that with $c=40^3$, the largest $c$ we performed in the test. Note that all power spectra shown in Figure \ref{fig_capacity_powspec} follow the $k^4$-minimal power spectrum.

Third, we adopt CCVTs with different capacities to perform a set of simulations in the cold dark matter (CDM) and warm dark matter (WDM) models. For CDM simulations, the set-ups are the same as those simulations with $256^3$ particles described in Section \ref{sec_app}. The WDM simulations only differ from the CDM ones in the input initial power spectra. Here, we adopt the fitted power spectrum given in \citet[][]{bode2001} with the parameter $g_X=1.5$, and set the mass for WDM particles as $m_X = 0.2$ keV. The thermal velocities for WDM particles are not included in our simulations. The two-point correlation functions and cumulative halo mass functions at $z=0$ from these simulations are plotted in Figure \ref{fig_capacity_sim}. As a comparison, we also plot the results from simulations with grid and glass pre-initial conditions.

In the CDM case, from Panels (a) and (b) of Figure \ref{fig_capacity_sim}, we find that with a lower capacity, the simulation tends to be artificially more clustered at small scales, and tends to artificially form more small structures. The fiducial results (i.e., the results from the simulation with $c=40^3$) agree with the grid/glass results at a level of $\la 2\%$, which indicates that a CCVT with $c=40^3$ has converged fairly well. For two-point correlation functions, in order to converge to the fiducial results on scales $r\geq \epsilon_{256^3}$ at a $\la 2\%$ ($\la 3\%$) level, we should adopt a capacity parameter $c \ga 20^3$ ($c \ga 10^3$). For cumulative halo mass functions, with a capacity of $c \ga 20^3$ ($c \ga 10^3$), the simulation can converge to the fiducial one at a level of $\la 1\%$ ($\la 2\%$) for halos with more than $10$ particles.

Similar conclusions can be drawn for WDM simulations (see Panels (c) and (d) of Figure \ref{fig_capacity_sim}). Especially, at $M_{200} \la 3 \times 10^{11}h^{-1}M_\odot$ (i.e., mass scales where the glass results differ from the grid ones at a level of $\ga 2\%$), the halo mass functions are affected by the discreteness effects \citep[see e.g.,][for detailed discussions]{wang2007} in all simulations. A simulation with a smaller capacity tends to experience stronger discreteness effects. To have a convergence level of $\la 2\%$ ($\la 3\%$) with respect to the fiducial simulation at $M_{200} \geq 3 \times 10^{11}h^{-1}M_\odot$, we should adopt a capacity of $c \ga 20^3$ ($c \ga 10^3$).

From the tests above, we conclude that a capacity of $c \geq 10^3$ should be used in cosmological simulations. In practice, to save computational time, we usually adopt a value between $c=10^3$ and $c=20^3$ in the current version of our CCVT code.

\section{Validity of CCVT tiles}\label{ap_tile}
\begin{figure} 
\includegraphics[width=245pt]{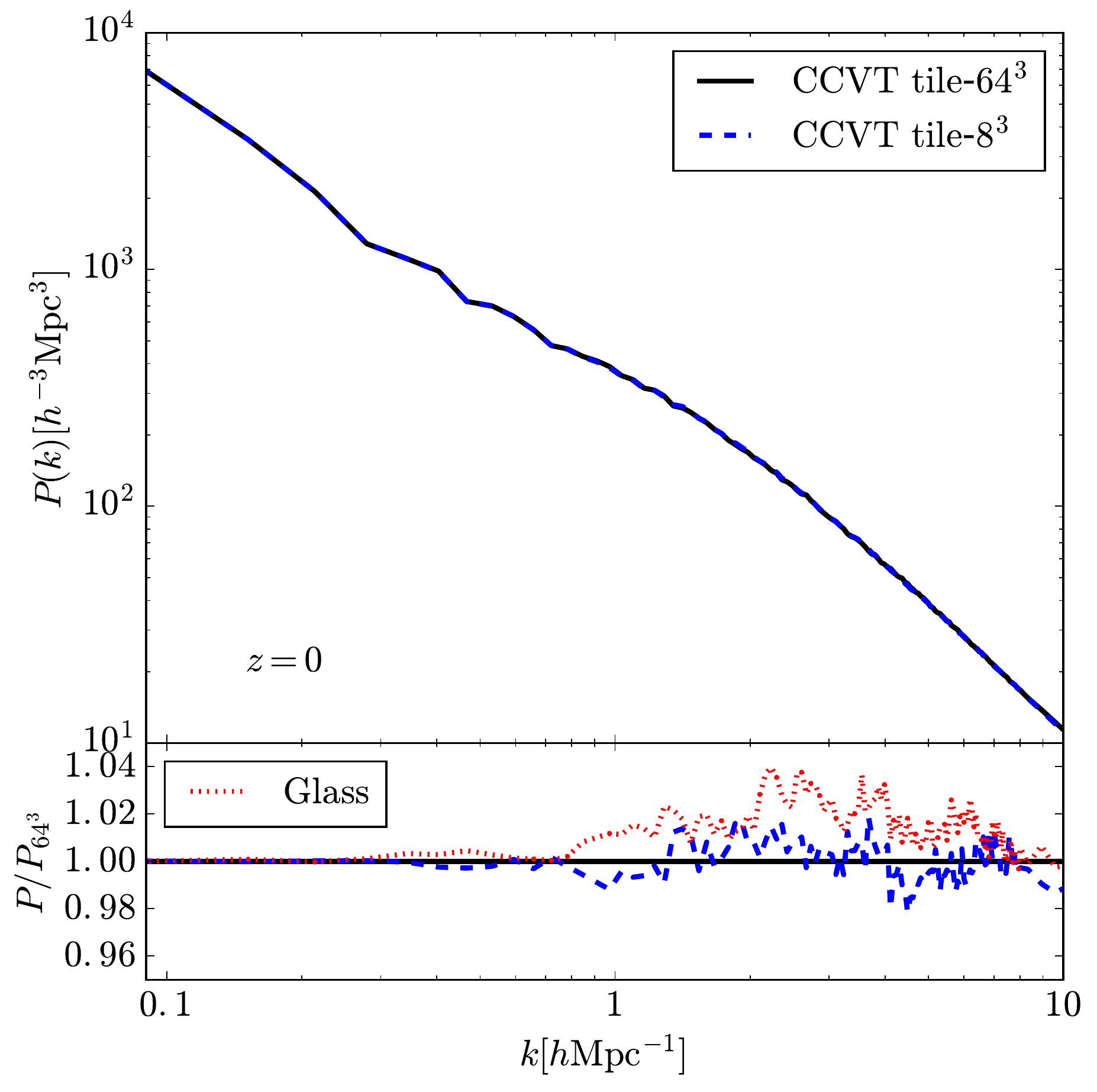} 
\caption{Power spectra of simulations from a CCVT of $64^3$ particles (black solid) and a CCVT of $8^3$ particles (blue dashed). The upper panel shows the power spectra at $z=0$, whereas the lower panel plots the ratio with respect to the power spectrum of the simulation with a $64^3$-CCVT, $P/P_{64^3}$. As a comparison, we have performed a similar set of simulations with glass pre-initial conditions. In the lower panel, we use the red dotted line to plot the ratio between the power spectrum of the simulation with tiling glass pre-initial conditions and that of the simulation without tiling glass pre-initial conditions.}\label{fig_tile_test}
\end{figure}

To overcome the computational challenge of making large CCVT pre-initial conditions, one solution is to get large CCVT configurations by tiling from small periodic CCVTs. Similar scheme is usually adopted when generating initial conditions from glass files. To validate the use of tiling CCVT pre-initial conditions, we perform two $\Lambda$CDM cosmological simulations with $64^3$ particles whose initial conditions are generated from a CCVT with $64^3$ particles and a CCVT with $8^3$ particles respectively. The former does not use tilings, while the latter uses $512$ tiles. Both simulations start at $z_\mathrm{IC}=40$ with the same cosmological parameters as the simulations mentioned in Section \ref{sec_app}. The box sizes are $100$ $h^{-1}\mathrm{Mpc}$, and the comoving softening lengths are $1/50$ of the mean inter-particle separation. Their power spectra at $z=0$ are plotted in Figure \ref{fig_tile_test}, with the black solid (blue dashed) line showing the former (latter) simulation. The difference in their power spectra is within $2\%$. As a comparison, we perform a parallel set of simulations with glass pre-initial conditions. In the lower panel of Figure \ref{fig_tile_test}, we use a red dotted line to plot the ratio between the power spectrum of the simulation starting with a glass initial condition tiled from $8^3$ particles and that of a simulation starting from a glass initial condition without tilings. The glass case shows similar results, i.e., the effects of tilings in power spectra are $\la 4\%$. Therefore, we conclude that it is valid to use the tiling scheme for CCVT pre-initial conditions.

\label{lastpage}

\end{document}